\documentclass{ptephy_v1-2007.08485v7}  %%FRK for v7 preprint on arXiv

\preprintnumber{arXiv:2007.08485}     %% Insert preprint number here

\usepackage{amsmath,amssymb,revsymb,graphicx,dcolumn}

\begin{document}

\title{IIB matrix model: Emergent spacetime from the master field}

\author{F.R. Klinkhamer$^\ast$}
\affil{Institute for Theoretical Physics,
Karlsruhe Institute of Technology (KIT),\\
76128 Karlsruhe, Germany \email{frans.klinkhamer@kit.edu}}

\begin{abstract}%
We argue that the large-$N$ master field
of the Lorentzian IIB matrix model can give the
points and metric of a classical spacetime.
\end{abstract}

\subjectindex{B25,\, B83 \vspace*{-7mm}}  

\maketitle

%%\newpage%%tmp
\section{Introduction}
\label{sec:Intro}

The IIB matrix model~\cite{IKKT-1997,Aoki-etal-review-1999}
has been suggested as a nonperturbative
formulation of type-IIB superstring theory. 
First results on the partition function %%XXX
of the Euclidean IIB matrix model
were reported in
Refs.~\cite{KrauthNicolaiStaudacher1998,AustingWheater2001}.
Later, numerical
simulations~\cite{KimNishimuraTsuchiya2012,NishimuraTsuchiya2019,%
Hatakeyama-etal2020}
of the Lorentzian IIB matrix model suggested the appearance of a $3+6$
split of the nine spatial dimensions
(matching Euclidean results were presented in
Ref.~\cite{Anagnostopoulos-etal-2020}).
Still, the physical interpretation of the emergence of
a classical spacetime in
Refs.~\cite{IKKT-1997,Aoki-etal-review-1999,KimNishimuraTsuchiya2012,%
NishimuraTsuchiya2019,Hatakeyama-etal2020,Anagnostopoulos-etal-2020}
is not really satisfactory, because there is
no manifest small dimensionless parameter
to motivate a saddle-point approximation.

Recently, we have revived an old idea, the large-$N$ master field
of Witten~\cite{Witten1979}, for a possible origin of classical spacetime
in the context of the IIB matrix model;
see App.~B in the earlier preprint
version~\cite{Klinkhamer2019-emergence-v6}
of Ref.~\cite{Klinkhamer2019-emergence-PTEP}.
But we did not give any details about where precisely in the master
field the classical spacetime is encoded. In the present paper,
we try to be more explicit.

Before we set out on our search for classical spacetime
in the IIB matrix model, we have five preliminary remarks.
First, we take the Lorentzian signature in the IIB matrix model,
because it is not clear how to interpret an emerging Euclidean  
``spacetime'' from the Euclidean IIB matrix model.
Second, our discussion of the Lorentzian path integrals will be strictly formal,
omitting all convergence issues.
Third, we introduce a length scale ``$\ell$'' into the IIB matrix model,
in order to give the dimension of length to the bosonic matrix variable.
Fourth, such a length scale ``$\ell$'' may enter the
effective metric of the regularized big bang
singularity~\cite{Klinkhamer2019,Klinkhamer2020,%
KlinkhamerWang2019-cosm,KlinkhamerWang2020-pert}.
Fifth, the focus of the present paper is solely
on the IIB matrix model, but it is possible that
some of our results could carry over to other matrix
models~\cite{deWitHoppeNicolai-1988,BanksFischlerShenkerSusskind-1997,%
DijkgraafVerlindeVerlinde-1997}.

We will now start by recalling the IIB matrix model
and the concept of the master field,
and will then turn to the emergence
of the spacetime points and the spacetime metric.

%%\newpage%%tmp
\section{Model}
\label{sec:Model}

The action of the Lorentzian IIB matrix model is given
by~\cite{IKKT-1997,Aoki-etal-review-1999}
\begin{subequations}\label{eq:IIB-matrix-model-action-etamunu}
\begin{eqnarray}
\label{eq:IIB-matrix-model-action}
S[A\,,\Psi] &=&
S_{b}[A]+S_{f}[A\,,\Psi]
\nonumber\\
&=&
\text{Tr}\, \Bigg(
\frac{1}{4}\,\big[ A^{\,\mu} ,\,A^{\nu}    \big]\,
             \big[ A^{\kappa},\,A^{\lambda} \big]\,
             \widetilde{\eta}_{\mu\kappa}\,\widetilde{\eta}_{\nu\lambda}
+\frac{1}{2}\, \overline{\Psi}_{\beta}\,
\widetilde{\Gamma}^{\,\mu}_{\beta\alpha}\,\,\widetilde{\eta}_{\mu\nu}
\big[ A^{\nu},\,\Psi_{\alpha} \big]\,
\Bigg)\,,
\\[2mm]
\label{eq:IIB-matrix-model-etamunu}
\widetilde{\eta}_{\mu\nu} &=&
\Big[ \text{diag}
\left(  -1,\,  1,\, \ldots \,,\,1 \right)
\Big]_{\mu\nu}\,,
\end{eqnarray} \end{subequations}
with vector indices
$\mu,\nu,\kappa,\lambda \in \{0,\,  1,\, \ldots \,,\,9\} $
and spinor indices $\alpha,\beta \in \{1,\,  2,\, \ldots \,,\,32\}$.
The vector $A^{\,\mu}$
and the Majorana--Weyl spinor $\Psi_{\alpha}$
are both $N \times N$ traceless Hermitian matrices. They
live in a 10D spacetime consisting of a single point,
a special case of the Eguchi--Kawai reduction~\cite{EguchiKawai1982}
operative in the large-$N$ limit of certain field theories;
see Ref.~\cite{Das1987} for a review.

The action \eqref{eq:IIB-matrix-model-action-etamunu}
is invariant under the following global gauge transformation:
\begin{subequations}\label{eq:IIB-matrix-model-global-gauge-transformation}
\begin{eqnarray}
\label{eq:IIB-matrix-model-global-gauge-transformation-A}
A^{\,\mu} &\to&  \Omega\, A^{\,\mu}\,\Omega^{\dagger}\,,
\\[2mm]
\label{eq:IIB-matrix-model-global-gauge-transformation-Psi}
\Psi_{\alpha} &\to&  \Omega\, \Psi_{\alpha}\,\Omega^{\dagger}\,,
\\[2mm]
\Omega &\in&  SU(N)\,.
\end{eqnarray}
\end{subequations}
In addition, there is $SO(1,\,9)$ Lorentz invariance and
an $\mathcal{N}=2$ supersymmetry~\cite{Aoki-etal-review-1999}.

The partition function $Z$ is defined by the following
Lorentzian ``path'' integral~\cite{KimNishimuraTsuchiya2012}:
\begin{eqnarray}
\label{eq:IIB-matrix-model-ZwithSgeneral}
Z &=&\int dA\,d\Psi\;\exp\left(i\,S[A\,,\Psi]/ \ell^{4}\,\right)\,.
\end{eqnarray}
Here, we have introduced a length scale ``$\ell$'', so that
$A^{\,\mu}$ from Eq.~\eqref{eq:IIB-matrix-model-action-etamunu}
must have the dimension of length and $\Psi_{\alpha}$
the dimension of $(\text{length})^{3/2}$.

The length scale ``$\ell$'' is solely introduced
to simplify the physics discussion later on and can
be removed by considering dimensionless variables
$A^{\prime}$ and $\Psi^{\prime}$.
The IIB-matrix-model path integral \eqref{eq:IIB-matrix-model-ZwithSgeneral}
in terms of dimensionless variables $A^{\prime}$
and $\Psi^{\prime}$ has, as emphasized
in App.~B of Ref.~\cite{Klinkhamer2019-emergence-v6},
no obvious small dimensionless parameter and, therefore,
no obvious saddle-point approximation.

As the fermions appear quadratically in the action, they
can be integrated
out~\cite{KrauthNicolaiStaudacher1998,AustingWheater2001}
and the partition function becomes
\begin{subequations}\label{eq:IIB-matrix-model-ZwithSeff-Seff}
\begin{eqnarray}
\label{eq:IIB-matrix-model-ZwithSeff}
Z &=&\int dA\;\exp\left(i\,S_\text{eff}[A]/ \ell^{4}\,\right) \,,
\end{eqnarray}
with an effective action
\begin{eqnarray}
\label{eq:IIB-matrix-model-Seff}
S_\text{eff}[A]&=&S_{b}[A]+S_\text{induced}[A]\,.
\end{eqnarray}
\end{subequations}
For completeness, we mention that the integration
measure $dA$ in Eqs.~\eqref{eq:IIB-matrix-model-ZwithSgeneral}
and \eqref{eq:IIB-matrix-model-ZwithSeff}
is standard~\cite{Brezin-etal-1979}, except
for the restriction to tracelessness.

%%\newpage%%tmp
\section{Master field}
\label{sec:Master-field}

A particular gauge-invariant bosonic observable
is given by
\begin{equation} \label{eq:IIB-matrix-model-w-observable}
w^{\mu_{1} \,\cdots\, \mu_{m}}  
=
\text{Tr}\,\big( A^{\mu_{1}} \cdots\, A^{\mu_{m}}\big)\,.
\end{equation}
Its expectation values are given by
the following  Lorentzian path integrals:
\begin{equation} \label{eq:IIB-matrix-model-w-product-vev}
\langle
w^{\mu_{1}\,\cdots\,\mu_{m}}\;w^{\nu_{1}\,\cdots\,\nu_{n}}\, \cdots\, \rangle
=
Z^{-1}\,\int dA\;
\big(w^{\mu_{1}\,\cdots\,\mu_{m}}\;w^{\nu_{1}\,\cdots\,\nu_{n}}\, \cdots\,\big)
\;\exp\left[i\,S_\text{eff}/ \ell^{4}\,\right]\,,
\end{equation}
with normalization factor $Z$ 
from Eq.~\eqref{eq:IIB-matrix-model-ZwithSeff-Seff}.

The expectation values \eqref{eq:IIB-matrix-model-w-product-vev}
have the following factorization property:
\begin{equation} \label{eq:IIB-matrix-model-w-product-factorization}
\langle w^{\mu_{1}\,\cdots\,\mu_{m}}\;
        w^{\nu_{1}\,\cdots\,\nu_{n}}\;
        \cdots \;
        w^{\omega_{1}\,\cdots\,\omega_{z}} \rangle
\stackrel{N}{=}
\langle w^{\mu_{1}\,\cdots\,\mu_{m}}\rangle\;
\langle w^{\nu_{1}\,\cdots\,\nu_{n}}\rangle\;
\cdots \;
\langle w^{\omega_{1}\,\cdots\,\omega_{z}} \rangle\,,
\end{equation}
which holds to leading order in $N$
(see Sect.~III A of Ref.~\cite{Das1987} for further discussion).
From Eq.~\eqref{eq:IIB-matrix-model-w-product-factorization}
follows the result that, to leading order in $N$,
the expectation value of the square of $w$ equals
the square of the expectation value of $w$,
\begin{equation}
\label{eq:IIB-matrix-w-square-vev}
\langle \, \big(w^{\mu_{1}\,\cdots\,\mu_{m}} \big)^{2} \, \rangle
\stackrel{N}{=}
\big(\,\langle w^{\mu_{1}\,\cdots\,\mu_{m}}  \rangle\, \big)^{2} \,,
\end{equation}
which is a truly remarkable result for a  statistical
(quantum) theory.

According to Witten~\cite{Witten1979}, the factorization results
\eqref{eq:IIB-matrix-model-w-product-factorization} and
\eqref{eq:IIB-matrix-w-square-vev} imply
that the path integrals \eqref{eq:IIB-matrix-model-w-product-vev}
are saturated by a \emph{single} configuration, the master field
$\widehat{A}^{\,\mu}$. For just one observable $w$
from Eq.~\eqref{eq:IIB-matrix-model-w-observable}
and its expectation value (``Wilson loop''), we then have
\begin{equation}
\label{eq:IIB-matrix-model-observable-from-master-field}
\langle w^{\mu_{1}\,\cdots\, \mu_{m}} \rangle
\stackrel{N}{=}
\text{Tr}\,\Big( \widehat{A}^{\,\mu_{1}} \cdots\, \widehat{A}^{\,\mu_{m}}\Big)\,.
\end{equation}
In principle, it is possible that there is more than one master field,
as long as these master fields give, in the large-$N$ limit,
exactly the same results for all possible observables of the 
type \eqref{eq:IIB-matrix-model-w-observable}.
For simplicity, we will talk, in the following,
about a single master field.

The explicit expression for the IIB-matrix-model
master field $\widehat{A}^{\,\mu}$ is not known, but 
it is possible to give an \emph{algebraic equation} for it.
Based on previous work by
Greensite and Halpern~\cite{GreensiteHalpern1983},
the IIB-matrix-model master field takes the following
form~\cite{Klinkhamer2019-emergence-v6}:
\begin{subequations}\label{eq:IIB-matrix-model-master-field-algebraic-equation}
\begin{eqnarray}\label{eq:IIB-matrix-model-master-field}
\widehat{A}^{\,\mu}_{\;ab}(\tau_\text{eq})
&=&
\exp\big[i\,(\widehat{p}_{a}-\widehat{p}_{b})\,\tau_\text{eq}\big]\;\,
\widehat{a}^{\,\mu}_{\;ab}\,,
\end{eqnarray}
where $\tau_\text{eq}$ must have a sufficiently large value  (it traces back
to the fictitious Langevin time $\tau$ of stochastic quantization)
and where the $\tau$-independent matrix $\widehat{a}^{\,\mu}$
on the right-hand side solves the following algebraic equation:
\begin{eqnarray}
\label{eq:IIB-matrix-model-algebraic-equation}
i\,\big(\widehat{p}_{a}-\widehat{p}_{b}\big)\;
\widehat{a}^{\,\mu}_{\;ab}
&=&
-\left.\frac{\delta S_\text{eff}}{\delta A_{\mu\;ba}}
 \right|_{A=\widehat{a}}\;
+\widehat{\eta}^{\,\mu}_{\;ab}\,,
\end{eqnarray}
\end{subequations}
in terms of the master momenta
$\widehat{p}_{a}$ (uniform random numbers)
and the master noise matrices $\widehat{\eta}^{\,\mu}_{\;ab}$
(Gaussian random numbers);
see Ref.~\cite{GreensiteHalpern1983} for further details
and  Refs.~\cite{Carlson-etal-1983,AlbertyGreensite1984} for some
interesting results.

Further remarks on the IIB-matrix-model master field also appear
in App.~B of Ref.~\cite{Klinkhamer2019-emergence-v6}, but, here,
we just assume that the master field has been obtained,
in the form as given
by Eq.~\eqref{eq:IIB-matrix-model-master-field-algebraic-equation}
or otherwise.

%%\newpage%%tmp
\section{Emergent spacetime points}
\label{sec:Emergent-spacetime-points}

As argued in App.~B of Ref.~\cite{Klinkhamer2019-emergence-v6},
the only place where ``classical spacetime'' can reside
in the IIB matrix model is
the master field $\widehat{A}^{\,\mu}$ of the model.
But precisely where? In the following, we present a few
rather naive ideas (hopefully, not too naive).

Following Refs.~\cite{KimNishimuraTsuchiya2012,%
NishimuraTsuchiya2019,Hatakeyama-etal2020}, we begin
by making a particular global gauge transformation
\eqref{eq:IIB-matrix-model-global-gauge-transformation-A},%
\begin{subequations}\label{eq:Hmu-bar-U0bar}
\begin{eqnarray} \label{eq:Hmu-bar}
\vspace*{-1mm}
\underline{\widehat{A}}^{\,\mu}
&=&
\underline{\Omega}\,\widehat{A}^{\,\mu}\,\underline{\Omega}^{\,\dagger}\,,
\\[2mm]
\label{eq:U0bar}
\underline{\Omega} &\in&  SU(N) \,,
\vspace*{-1mm}
\end{eqnarray}
\end{subequations}
so that the transformed 0-component [singled out by the Minkowski
``metric'' \eqref{eq:IIB-matrix-model-etamunu}]
is diagonal and has ordered eigenvalues
$\widehat{\alpha}_{i} \in \mathbb{R}$,
\begin{subequations}\label{eq:H0hat-bar-alphahat-order}
\begin{eqnarray} \label{eq:H0hat-bar}
\vspace*{-1mm}
\underline{\widehat{A}}^{\,0}
&=&
\text{diag} \Big( \widehat{\alpha}_{1},\,\widehat{\alpha}_{2},
\,\ldots\,,\,
\widehat{\alpha}_{N-1},\,\widehat{\alpha}_{N} \Big)\,,
\\[2mm]
\label{eq:alphahat-order}
\widehat{\alpha}_{1}
&\leq&
\widehat{\alpha}_{2}\,\leq\,\;\cdots\;\,\leq\,
\widehat{\alpha}_{N-1}\,\leq\,\widehat{\alpha}_{N}\,,
\\[2mm]
\label{eq:alphahat-sum-zero}
\sum_{i=1}^{N}\,\widehat{\alpha}_{i}&=& 0\,,
\vspace*{-1mm}
\end{eqnarray}
\end{subequations}
where the last equality from tracelessness implies that 
some $\widehat{\alpha}_{i}$ are negative and some positive.  
The ordering \eqref{eq:alphahat-order} will turn out to be crucial for
the time coordinates $\widetilde{t}$ and $\widehat{t}$ obtained below.

Indeed, we 
can  %%XXX
introduce a continuous function
$\widetilde{x}^{\,0}\,(\widetilde{\zeta})
\equiv \widetilde{c}\;\widetilde{t}\,(\widetilde{\zeta})$
for $\widetilde{\zeta}\in (0,\,1]$ by identifying
(cf. Ref.~\cite{Brezin-etal-1979})%
\begin{equation} \label{eq:t-tilde-def}
\vspace*{-1mm}
\widetilde{x}^{\,0}\,(i/N)
\equiv
\widetilde{c}\;\widetilde{t}\,(i/N)
=
\widehat{\alpha}_{i}\,,
\vspace*{-1mm}
\end{equation}
with $i \in \{1,\,\ldots ,\,  N\}$
and a velocity $\widetilde{c}$  that is expected to be related to the
vacuum velocity of light in the low-energy theory.
From Eq.~\eqref{eq:alphahat-order}, we immediately have
\begin{equation}
\label{eq:ttilde-order}
\vspace*{-1mm}
\widetilde{t}\,\big(1/N\big) \,\leq\, \widetilde{t}\,\big(2/N\big)\,\leq
\;\cdots\;
\leq\, \widetilde{t}\,\big(1-1/N\big)\,\leq\, \widetilde{t}\,\big(1\big)\,,
\vspace*{-1mm}
\end{equation}
where the ordering is the defining property of what makes physical time.

The problem now is how to extract the \emph{corresponding}
space coordinates $\widetilde{x}^{\,m}(\widetilde{\zeta})$
from the Hermitian $\underline{\widehat{A}}^{\,m}$ matrices.  
The simplest idea (following Ref.~\cite{Aoki-etal-review-1999})
is to calculate the eigenvalues of the nine matrices
$\underline{\widehat{A}}^{\,m}$, but then 
it is unclear how to order them with respect to the eigenvalues
from Eq.~\eqref{eq:H0hat-bar-alphahat-order}.
We will use a relatively simple procedure, which approximates
the $\underline{\widehat{A}}^{\,m}$ eigenvalues but still manages
to order them along the diagonal. Our procedure corresponds, in fact,
to a type of coarse graining of some of the
information contained in the IIB-matrix-model master field.
There is, however, more information in the master field that we will
not consider, and even information not in the master field, as there
are also non-factorizing observables~\cite{Das1987}
in the IIB matrix model.

We start from the following trivial observation:
if $M$ is an $N\times N$ Hermitian matrix, then \emph{any}
$n\times n$ block centered on the diagonal of $M$
is \emph{also} Hermitian, which holds for $n\geq 1$ and $n \leq N$.
With $N \gg1$, we take $n$ so that $1 \ll n \ll N$.
Specifically, we proceed by the following six steps.

%%\newpage%%tmp
The first step is to let $K$ be an odd divisor of $N$, so that
\begin{subequations}\label{eq:N-as-product-K-times-n}
\begin{eqnarray}
\vspace*{-1mm}
N&=&K\,n\,,
\\[1mm]
K &=& 2\,L+1\,,
\vspace*{-1mm}
\end{eqnarray}
\end{subequations}
where both $L$ and $n$ are positive integers (we have chosen an
odd value of $K$ for later convenience).
In the limit $N\to\infty$, we also take $K\to\infty$
but are not sure exactly how fast (with $n$ staying finite or not).

The second step is to consider, 
in each of the ten matrices $\underline{\widehat{A}}^{\,\mu}$ from 
Eqs.~\eqref{eq:Hmu-bar-U0bar} and \eqref{eq:H0hat-bar-alphahat-order},
the $K$ blocks of size $n\times n$ centered on the diagonals.

The third step is to realize that we already know the diagonalized
blocks of $\underline{\widehat{A}}^{\,0}$ from
Eq.~\eqref{eq:H0hat-bar}. This allows us to
define the following time coordinate
$\widehat{t}\,(\zeta)$, for $\zeta\in (0,\,1]$,
as the average of the $\widehat{\alpha}_{i}$
eigenvalues of each $n\times n$ block:
\begin{equation} \label{eq:x0hat-def}
\vspace*{-1mm}
\widehat{x}^{\,0}\,\big(k/K\big) \equiv
\widetilde{c}\;\widehat{t}\,\big(k/K\big) \equiv
\left(\frac{1}{n}\;\sum_{j=1}^{n} \, \widehat{\alpha}_{(k-1)\,n+j}\right)
+ \widetilde{c}\;\widehat{t}_\text{shift}\,,
\vspace*{-1mm}
\end{equation}
with  $k \in \{1,\,\ldots ,\,  K\}$, an arbitrary real constant
$\widehat{t}_\text{shift}$, and the
velocity $\widetilde{c}$ mentioned below Eq.~\eqref{eq:t-tilde-def}.
The time coordinates from Eq.~\eqref{eq:x0hat-def} are ordered,
\begin{equation}
\label{eq:that-order}
\vspace*{-1mm}
\widehat{t}\,\big(1/K\big) \,\leq\, \widehat{t}\,\big(2/K\big)\,\leq
\;\cdots\;
\leq\, \widehat{t}\,\big(1-1/K\big)\,\leq\, \widehat{t}\,\big(1\big)\,,
\vspace*{-1mm}
\end{equation}
because the $\widehat{\alpha}_{i}$ are, according to
Eq.~\eqref{eq:alphahat-order}.
With an appropriate value of %%XXX
$\widehat{t}_\text{shift}$ in Eq.~\eqref{eq:x0hat-def},
we can set $\widehat{t}=0$ for the halfway block at $k=L+1$.
The blocks with $k<L+1$ will generically have
negative time coordinates $\widehat{t}$ and
those with $k>L+1$
\mbox{generically positive time coordinates $\widehat{t}$.}

The fourth step is to
obtain the eigenvalues of the $n\times n$ blocks of the nine
spatial matrices $\underline{\widehat{A}}^{\,m}$
and to denote these real eigenvalues $\big(\widehat{\beta}^{\,m}\big)_{i}\,$,
with $i \in \{1,\,\ldots ,\,  N\}$.
How the $n$ eigenvalues are ordered in each block is irrelevant,
as they will be averaged over in the next step.

The fifth step is to define, just as in step three,
the following nine spatial coordinates $\widehat{x}^{\,m}(\zeta)$,
for $\zeta\in (0,\,1]$, as the averages of the
$\big(\widehat{\beta}^{\,m}\big)_{i}\,$ eigenvalues of the $n\times n$ blocks:
\begin{equation} \label{eq:xmhat-def}
\vspace*{-1mm}
\widehat{x}^{\,m}\big(k/K\big)
\equiv
\frac{1}{n}\;\sum_{j=1}^{n}\, \left[\,\widehat{\beta}^{\,m}\,\right]_{(k-1)\,n+j}\,,
\vspace*{-1mm}
\end{equation}
with  $k \in \{1,\,\ldots ,\,  K\}$. The averaging is done independently
for each value of $m$.

The sixth and last step is, first, to observe that
$\widehat{t}\,(\zeta)$ 
from Eqs.~\eqref{eq:x0hat-def} and \eqref{eq:that-order}
is a nondecreasing function of $\zeta\equiv k/K$ and, then,
to eliminate $\zeta$ between $\widehat{t}\,(\zeta)$ 
from Eq.~\eqref{eq:x0hat-def}
and $\widehat{x}^{\,m}(\zeta)$ from Eq.~\eqref{eq:xmhat-def}, 
in order to obtain
\begin{equation} \label{eq:xm-from-t}
\widehat{x}^{\,m}=\widehat{x}^{\,m}\big(\,\widehat{t}\;\big)\,,
\end{equation}
which corresponds to a particular foliation of what will become
the classical spacetime.

%%\newpage%%tmp
If the master-field matrices $\underline{\widehat{A}}^{\,\mu}$
are more or less block-diagonal (with a width $\Delta N \ll N$,
as suggested by the numerical results from
Refs.~\cite{KimNishimuraTsuchiya2012,NishimuraTsuchiya2019,Hatakeyama-etal2020})
and if an appropriate value of $n$
can be chosen (perhaps $n \sim \Delta N$,
for sufficiently large values of $N$), then the
expressions \eqref{eq:x0hat-def} and \eqref{eq:xmhat-def}
may provide suitable spacetime points.
In a somewhat different notation, these spacetime points are denoted
\begin{equation}
\label{eq:xhat-mu-k}
\vspace*{-1mm}
\widehat{x}^{\,\mu}_{k}=
\left(\,\widehat{x}^{\,0}_{k},\, \widehat{x}^{\,m}_{k}\,\right)
\equiv
\Big(\,\widehat{x}^{\,0}\big(k/K\big),\,
\widehat{x}^{\,m}\big(k/K\big)\,\Big)\,,
\vspace*{-1mm}
\end{equation}
where $k$ runs over $\{1,\,\ldots ,\,  K\}$
with $K$ given by Eq.~\eqref{eq:N-as-product-K-times-n}.
Each of these ten coordinates has the dimension of length,
which traces back to the dimension of the bosonic matrix
variable $A^{\,\mu}$, as discussed in Sect.~\ref{sec:Model}.
The points \eqref{eq:xhat-mu-k}, and those obtained from
different choices of block size $n$
and block position along the diagonals of
the master-field matrices,
effectively build a spacetime manifold
with continuous (interpolating) coordinates $x^{\,\mu}$
if there is also an emerging metric $g_{\mu\nu}(x)$.

%%\newpage%%tmp
\section{Emergent spacetime metric}
\label{sec:Emergent-spacetime-metric}

In Sect.~\ref{sec:Emergent-spacetime-points}, we have
obtained $K$ points $\widehat{x}^{\,\mu}_{k}$
as given by Eq.~\eqref{eq:xhat-mu-k},
which sample a 10D classical spacetime.
(We have put a hat on our coordinates, in order to
remind us of their master-field origin.)
The idea now is that low-energy fields 
propagate over a spacetime manifold which interpolates
between these discrete spacetime points $\widehat{x}^{\,\mu}_{k}$.
The low-energy fields include the matter fields (scalar,
vector, spinor) and the metric field (tensor). In fact,
Aoki et al.~\cite{Aoki-etal-review-1999}
have argued that the propagation of a matter field
(for example, the propagation of a scalar field $\sigma$)
determines the effective inverse metric, which is found to depend on
the density function of the spacetime points $\widehat{x}^{\,\mu}_{k}$
and the correlations of these density functions.

The crucial result in Ref.~\cite{Aoki-etal-review-1999}
is Eq.~(4.16), which we rewrite as follows:
\begin{equation} \label{eq:emergent-inverse-metric}
g^{\mu\nu}(x) \sim
\int_{\mathbb{R}^{D}} d^{D}y\;
\langle\langle\, \rho(y)  \,\rangle\rangle
\; (x-y)^{\,\mu}\,(x-y)^{\nu}\;f(x-y)\;r(x,\,y)\,,
\end{equation}
where $D=10$ is the spacetime dimension and the
average $\langle\langle\, \rho(y)  \,\rangle\rangle$ corresponds,
for the procedure used in Sect.~\ref{sec:Emergent-spacetime-points},
to averaging over different block sizes
and block positions along the diagonals of the master-field matrices
(details will be presented elsewhere).

The quantities that enter the multiple
integral \eqref{eq:emergent-inverse-metric}
are the density function
\begin{equation} \label{eq:rho-def}
\rho(x) \;\equiv \;
\sum_{k=1}^{K}\;\delta^{(D)} \big(x- \widehat{x}_{k}\big)\,,
\end{equation}
the dimensionless density correlation function $r(x,\,y)$ defined by
\begin{equation} \label{eq:r-def}
\langle\langle\,\rho(x)\,\rho(y) \,\rangle\rangle \;\equiv \;
\langle\langle\, \rho(x)\,\rangle\rangle\; \langle\langle\,\rho(y) \,\rangle\rangle\;
r(x,\,y)\,,
\end{equation}
and a strongly localized function $f(x)$,
which appears in the effective action of a low-energy scalar
degree of freedom $\sigma$  ``propagating'' over the discrete
spacetime points $\widehat{x}^{\,\mu}_{k}$,
\begin{equation} \label{eq:Seff-phi}
S_\text{eff}[\sigma] \propto
\sum_{k,\,l}\; \frac{1}{2}\,f\big(\widehat{x}_{k}-\widehat{x}_{l}\big)\;
\big( \sigma_{k}- \sigma_{l}  \big)^{2}
+ \sum_{k}\;\frac{1}{2}\,\mu^{2}\,\ell^{-2}\; \big( \sigma_{k}\big)^{2}\,,
\end{equation}
where $f(x)=f\left(x^{0},\,  x^{1},\, \ldots \,,\,x^{D-1}\right)$
has dimension $1/(\text{length})^{2}$,
$\mu$ is dimensionless, and $\ell$ is the model
length scale introduced
in Eq.~\eqref{eq:IIB-matrix-model-ZwithSgeneral}.
Here, $\sigma_{k}$ is the field value at the point
$\widehat{x}_{k}$ and the continuous field $\sigma(x)$
has $\sigma(\widehat{x}_{k})=\sigma_{k}\,$.
After averaging over different block structures
in the master-field matrices
(see above) and making a Taylor expansion,
the continuous field $\sigma(x)$ is found to have a standard
kinetic term
$g^{\mu\nu}\,\partial_{\mu} \sigma\,\partial_{\nu}\sigma$
in the action, with the inverse metric given by
Eq.~\eqref{eq:emergent-inverse-metric}. See Sect.~4.2 of %%XXX
Ref.~\cite{Aoki-etal-review-1999} for further details,
App.~\ref{app:Effective-action-scalar} for a sample calculation,
and Ref.~\cite{ChristFriedbergLee1982} for earlier work
on random-lattice field theories.

%%%%%%%%%%%%%%%%%%%%%%%\newpage%%tmp
The inverse metric $g^{\mu\nu}(x)$
from Eq.~\eqref{eq:emergent-inverse-metric} is manifestly dimensionless
and the metric $g_{\mu\nu}$ is simply obtained as the matrix inverse
of $g^{\mu\nu}$.
In fact,  general covariance is also expected to
emerge dynamically~\cite{Aoki-etal-review-1999} and
the quantity determined by the integral \eqref{eq:emergent-inverse-metric}
will, for a strongly localized function $f$,
transform approximately like $dx^{\mu}\,dx^{\nu}$,
that is, approximately like  %%XXX
a rank-2 contravariant tensor.
Taking the matrix inverse of this quantity gives an object
that transforms approximately like a rank-2 covariant tensor,
so that this object can indeed be interpreted as
the emergent metric $g_{\mu\nu}(x)$.

The outstanding tasks are
to obtain the master-field matrices $\widehat{A}^{\,\mu}$,
to identify an effective scalar $\sigma$ from it
(cf. Sect.~4.1 of Ref.~\cite{Aoki-etal-review-1999}),
and to recover the effective action \eqref{eq:Seff-phi}.
The explicit results for $\rho(x)$, $f(x)$, and $r(x,\,y)$  must also
explain how the inverse metric 
from Eq.~\eqref{eq:emergent-inverse-metric}
acquires a Lorentzian signature.

Using appropriate units %%XXX
to set $\ell =1$,
we have performed a toy-model calculation with the function
$f_\text{test,2}(x)=\alpha + x^{0} \, x^{1}$ inserted into the
multiple integral \eqref{eq:emergent-inverse-metric} for $D=2$,
where we also
assume $\rho(x)=r(x,\,y)=1$ and cut the integration ranges off
symmetrically at $\pm 1$.
The resulting inverse metric at $x^{\mu}=0$ is found to
change continuously from a Euclidean to a Lorentzian signature as the
parameter $\alpha$ changes continuously from $\alpha=1$ to $\alpha=0$
(see App.~\ref{app:Emergent-Lorentzian-signature} for further details
and a trivial extension to $D=4$).
The conclusion is that,
in principle, it is possible to obtain a Lorentzian inverse metric
from the expression \eqref{eq:emergent-inverse-metric}.
But  %%XXX
it will be a challenge to establish, if at all relevant,
the effective Lorentzian metric of the regularized big bang
singularity with $b\sim \ell$ as the 
length parameter~\cite{Klinkhamer2019,Klinkhamer2020,%
KlinkhamerWang2019-cosm,KlinkhamerWang2020-pert}.

For the record, we give
a further result, based on Eq.~(4.17) of Ref.~\cite{Aoki-etal-review-1999},
which concerns the background value of the dilaton field $\Phi$,
\begin{equation} \label{eq:emergent-dilaton}
\sqrt{-g(x)}\;\exp\big[- \Phi(x) \big]
\propto
\langle\langle\, \rho(x)  \,\rangle\rangle\,,
\end{equation}
with $g \equiv \det g_{\mu\nu}$ and the meaning of the
average on the right-hand side explained in the text below  
Eq.~\eqref{eq:emergent-inverse-metric}.

Returning to the expression \eqref{eq:emergent-inverse-metric}
for the emergent inverse metric,
we observe that it depends not only on the density distribution $\rho$
of emerged spacetime points and their correlation function $r$, but also on the
localization function $f$ from the scalar effective action \eqref{eq:Seff-phi}.
In this way, the metric only exists if matter is present, which
reminds us of Dicke's interpretation of spacetime
(see App. 4, p. 50 and App. 5, p. 60 in Ref.~\cite{Dicke1964}).
The new insight from the IIB matrix model is that matter and
spacetime are expected to emerge simultaneously.

\section*{Note added}

Two subsequent papers~\cite{Klinkhamer2020b,Klinkhamer2020c}
give details on the extraction
of the spacetime points and the spacetime metric, assuming 
that the IIB-matrix-model master field is known.
A further paper~\cite{Klinkhamer2020d}
shows that the IIB-matrix-model master field can, in principle,
give rise to the regularized big bang metric~\cite{Klinkhamer2019}
of general relativity.

\section*{Acknowledgments}

It is a pleasure to thank J.~Nishimura and H.C. Steinacker
for comments on an earlier version of this article.
The referee is thanked for constructive remarks.

\appendix

%\section{Effective action for a scalar degree of freedom}
\section{Effective action of a scalar degree of freedom} %%XXX
\label{app:Effective-action-scalar}

The expression \eqref{eq:emergent-inverse-metric}
for the emergent inverse metric
in Sect.~\ref{sec:Emergent-spacetime-metric}
was obtained from an \emph{assumed}
effective action \eqref{eq:Seff-phi}
of a scalar degree of freedom $\sigma$.
Even though the particular form of this effective action
is entirely reasonable
(cf. the discussion of random-lattice scalars in Sect.~6
of Ref.~\cite{ChristFriedbergLee1982}), it 
is desirable to understand in some detail how
this effective action could arise in the IIB matrix model.
This is done in the present appendix,
where we show that the IIB matrix model
can, in principle,
produce the effective action \eqref{eq:Seff-phi}.

We start by noting that we should not be led astray by the
notation $A^{\,\mu}$ resembling ten gauge fields and that the
IIB-matrix-model master field $\widehat{A}^{\,\mu}$ is really
a \emph{single} $10\times N\times N$ matrix with entries having
the dimension of length.
The last observation suggests that, in order to get an
effective \emph{field} $\phi(x^{0},\, \ldots\, ,\,x^9)$
in the continuum, the perturbation $\phi_{k}$
of the master-field matrix must be taken  \emph{equal}
on all ten ``slices'' of the matrix
(an explicit example will be given below).

For simplicity, we focus on the four ``large'' spacetime
dimensions~\cite{KimNishimuraTsuchiya2012,NishimuraTsuchiya2019},
\begin{equation} %\label{eq:}
D=4\,,
\end{equation}
and 
let %%XXX
the indices $\mu,\,\nu, \ldots$ run over \{0,\,  1,\, 2,\,\,3\}.
We now present an explicit construction of a perturbation
of the master field for the case
\begin{equation}
\label{eq:N-n}
N=K\,n = 6\,,\quad  n=3\,,
\end{equation}
where $n$ corresponds to the averaging block used in
Sect.~\ref{sec:Emergent-spacetime-points}
for the extraction of the spacetime points
(here, there are only two spacetime points,
$\widehat{x}^{\,\mu}_{1}$ and $\widehat{x}^{\,\mu}_{2}$).
For the sake of argument, we simply assume that $N=6$ is large
enough, so that there exists a master field (later, we will
extend the explicit construction to $N\gg1$).

The $6\times 6$ master-field matrices are assumed to have a band-diagonal
structure~\cite{KimNishimuraTsuchiya2012,NishimuraTsuchiya2019,%
Hatakeyama-etal2020} and are given by
\begin{subequations}\label{eq:approx-master-field-matrices-N-is-6}
\begin{equation}
\underline{\widehat{A}}^{\,\mu}
=
\left(
  \begin{array}{cc}
\;\;\mathcal{B}^{\,\mu}_{11}\;\; &  \;\;\mathcal{B}^{\,\mu}_{12}\;\;\\[1mm]
\;\;\mathcal{B}^{\,\mu}_{21}\;\; &  \;\;\mathcal{B}^{\,\mu}_{22}\;\;\\
  \end{array}
\right)
\,,
\end{equation}
in terms of $3\times 3$ blocks $\mathcal{B}^{\,\mu}_{kl}$, where
\begin{equation}
\mathcal{B}^{\,\mu}_{12} \sim 0\,,
\quad
\mathcal{B}^{\,\mu}_{21}\sim 0\,,
\end{equation}
and the block $\mathcal{B}^{\,\mu}_{11}$ has real
eigenvalues $\{\widehat{x}^{\,\mu}_{1,a},\,\widehat{x}^{\,\mu}_{1,b},\,
\widehat{x}^{\,\mu}_{1,c}\}$ with an average value
\begin{equation}
%\label{eq:}
\widehat{x}^{\,\mu}_{1}=
\frac{1}{3}\,\left(
\widehat{x}^{\,\mu}_{1,a}+
\widehat{x}^{\,\mu}_{1,b}+
\widehat{x}^{\,\mu}_{1,c} \right)\,,
\end{equation}
and similarly for the block $\mathcal{B}^{\,\mu}_{22}$, with real
eigenvalues $\{\widehat{x}^{\,\mu}_{2,a},\,\widehat{x}^{\,\mu}_{2,b},\,
\widehat{x}^{\,\mu}_{2,c}\}$ and an average value
\begin{equation}
%\label{eq:}
\widehat{x}^{\,\mu}_{2}=
\frac{1}{3}\,\left(
\widehat{x}^{\,\mu}_{2,a}+
\widehat{x}^{\,\mu}_{2,b}+
\widehat{x}^{\,\mu}_{2,c} \right)\,.
\end{equation}
\end{subequations}

Now consider the following $6\times 6$ matrices
$A^{\,\mu}$ involving the
perturbations $\phi_{1},\,\phi_{2}$ $\in$ $\mathbb{R}$\,:%
\begin{subequations}\label{eq:perturbed-matrices-N-is-6}
\begin{equation}
\label{eq:perturbed-matrices-N-is-6-matrices}
A^{\,\mu} =
\text{diag}\,
\big(B^{\,\mu}_{<11>},\,B^{\,\mu}_{<12>},\, B^{\,\mu}_{<22>}\big)\,,
\end{equation}
in terms of $2\times 2$ blocks
\begin{eqnarray}
\label{eq:perturbed-matrices-N-is-6-block-left}
B^{\,\mu}_{<11>}
&=&  \left(
       \begin{array}{cc}
        \widehat{x}^{\,\mu}_{1}  &
        \;\;c^{\,\mu}\,\phi_{1}\,\left(1-\phi_{1}^{2}/\ell^{2}\right) \\
        c^{\,\mu}\,\phi_{1}\,\left(1-\phi_{1}^{2}/\ell^{2}\right)\;\;  &
        \widehat{x}^{\,\mu}_{1}+\phi_{1}\\
       \end{array}
     \right)\,,
\\[1.0mm]
\label{eq:perturbed-matrices-N-is-6-block-mid}
B^{\,\mu}_{<12>}
&=&  \left(
       \begin{array}{cc}
        \widehat{x}^{\,\mu}_{1}  &
        \;\;k_{12}\,\left(\phi_{1}-\phi_{2}\right) \\
        k_{12}\,\left(\phi_{1}-\phi_{2}\right)\;\; &
         \widehat{x}^{\,\mu}_{2} \\
       \end{array}
     \right)\,,
\\[1.0mm]
\label{eq:perturbed-matrices-N-is-6-block-right}
B^{\,\mu}_{<22>}
&=&  \left(
       \begin{array}{cc}
       \widehat{x}^{\,\mu}_{2}   &
      \;\; d^{\,\mu}\,\phi_{2}\,\left(1-\phi_{2}^{2}/\ell^{2}\right)\\
       d^{\,\mu}\,\phi_{2}\,\left(1-\phi_{2}^{2}/\ell^{2}\right)\;\;  &
        \widehat{x}^{\,\mu}_{2}+\phi_{2} \\
       \end{array}
     \right)\,,
\end{eqnarray}
for a dimensionless coupling $k_{12}$ and
dimensionless constants $c^{\,\mu}$ and $d^{\,\mu}$,
\begin{eqnarray}
k_{12}&=& k_{12}\big(\Delta x\big) \in \mathbb{R} \,,
\\[1.0mm]
c^{\,\mu}&=&
\left(  c^{0},\, c,\, c,\, c\right) \in \mathbb{R}^{4}\,,
\\[1.0mm]
d^{\,\mu}&=&
\left(  d^{0},\, d,\, d,\, d\right) \in \mathbb{R}^{4}\,,
\end{eqnarray}
\end{subequations}
with definition
\begin{eqnarray}
\Delta x^{\,\mu}         &\equiv&
\widehat{x}^{\,\mu}_{2} -\widehat{x}^{\,\mu}_{1} \,.
\end{eqnarray}
Three remarks are in order.
First, the \emph{same} perturbation $\phi_{1}$
appears in \emph{all} four matrices $A^{\,\mu}$,
and similarly for $\phi_{2}$.
Second, the parameter $k_{12}$ depends on the
coordinate distance $\Delta x^{\,\mu}$
and is assumed to drop rapidly as this distance increases
(otherwise, the emerging scalar theory does
not make sense~\cite{ChristFriedbergLee1982}).
Third, the matrices $A^{\,\mu}$ reduce,
for $\phi_{1}=\phi_{2}=0$,
to diagonal matrices with approximately the same
eigenvalues as the master-field
matrices \eqref{eq:approx-master-field-matrices-N-is-6},
which were assumed to be band-diagonal.

Next, insert the perturbation matrices $A^{\,\mu}$ from
Eq.~\eqref{eq:perturbed-matrices-N-is-6}
in the bosonic action \eqref{eq:IIB-matrix-model-action-etamunu}
and find%
\begin{eqnarray}\label{eq:Sb-phi-perturbations-N-is-6}
S_b\,\Big|^\text{(pert)} &=&
\frac{1}{2}\,\Big[
3\,\left(\Delta x^{0}\right)^{2}
-\left(\Delta x^{1}\right)^{2}
-\left(\Delta x^{3}\right)^{2}
-\left(\Delta x^{1}\right)^{2}
- 2\,\Delta x^{0}\,\left( \Delta x^{1} + \Delta x^{2} +  \Delta x^{3} \right) %
\nonumber\\[1mm]
&&
+ 2\,\Delta x^{1}\,\Delta x^{2}
+ 2\,\Delta x^{2}\,\Delta x^{3}
+ 2\,\Delta x^{3}\,\Delta x^{1}
\Big]\;
\Big(k_{12}\big(\Delta x\big)\Big)^{2}\, {\left( \phi_{1} - \phi_{2} \right) }^{2}
\nonumber\\[1mm]
&&
+\frac{2}{3}\,{\ell}^{-4}\,{(c^{0}-c)}^{2}\,{\phi_{1}}^{4}\,
       {\left( {\ell}^{2} - {\phi_{1}}^{2} \right) }^{2}
+ \frac{2}{3}\,{\ell}^{-4}\,{(d^{0}-d)}^{2}\,{\phi_{2}}^{4}\,
       {\left( {\ell}^{2} - {\phi_{2}}^{2} \right) }^{2}\,.
\end{eqnarray}
Apparently, we have already recovered the ``kinetic'' term
$\left(\sigma_{1}-\sigma_{2}\right)^{2}$
of Eq.~\eqref{eq:Seff-phi}, which gives rise to the
emergent inverse metric \eqref{eq:emergent-inverse-metric}.
The mass-squared terms ${\sigma_{1}}^{2}$ and ${\sigma_{2}}^{2}$
of Eq.~\eqref{eq:Seff-phi} result from spontaneous symmetry breaking,
at least for the simple model considered.
Indeed, with shifted scalar variables,
\begin{equation}
\label{eq:shifted-scalar-variables}
\phi_{1} =\ell+\chi_{1}\,,
\quad
\phi_{2} =\ell+\chi_{2}\,,
\end{equation}
the effective action \eqref{eq:Sb-phi-perturbations-N-is-6} becomes, 
in a shorthand notation, 
\begin{eqnarray}
\label{eq:Sb-chi-perturbations-N-is-6}
\hspace*{-0mm}
S_b\,\Big|^\text{(pert)} &=&
\frac{1}{2}\,
\Big[ \cdots \Big]\;
\Big(k_{12}\big(\Delta x\big)\Big)^{2}\, {\big( \chi_{1} - \chi_{2} \big) }^{2}
\nonumber\\[1mm]&&
+\, 6\,\big(c^{0}-c\big)^{2}\,\ell^{2}\,{\chi_{1}}^{2}
+ 6\,\big(d^{0}-d\big)^{2}\,\ell^{2}\,{\chi_{2}}^{2} + \cdots\,,
\end{eqnarray}
where the ellipsis at the end stands for
cubic and higher-order self-interaction
terms of the scalars $\chi_{1}$ and $\chi_{2}$. 
Note that the square bracket in
Eq.~\eqref{eq:Sb-chi-perturbations-N-is-6}, which is explicitly shown in
Eq.~\eqref{eq:Sb-phi-perturbations-N-is-6},
can be positive, zero, or negative, whereas the mass-square terms 
in Eq.~\eqref{eq:Sb-chi-perturbations-N-is-6} are strictly nonnegative.
The indefinite sign of the square bracket
in Eqs.~\eqref{eq:Sb-phi-perturbations-N-is-6}
and \eqref{eq:Sb-chi-perturbations-N-is-6}
traces back to the Lorentzian ``signature'' of
the coupling constants \eqref{eq:IIB-matrix-model-etamunu}
in the IIB matrix model.

By adding appropriate (generalized) blocks
to Eq.~\eqref{eq:perturbed-matrices-N-is-6-matrices}
we can easily obtain matrices with larger values of $N$.
In this way, we keep essentially the same properties as discussed
for the $(N,\,n)=(6,\,3)$ case and obtain, in particular,
an effective action with kinetic terms 
as shown Eq.~\eqref{eq:Seff-phi},
but now in terms of scalars $\chi_k$.

%%\newpage%%tmp
\section{Emergent Lorentzian signature}
\label{app:Emergent-Lorentzian-signature}

In this appendix, we present some details of the 
2D toy-model calculation for the emergent inverse metric
mentioned in Sect.~\ref{sec:Emergent-spacetime-metric}.
The aim of this 2D toy-model calculation is to present
a possible mechanism for obtaining, in the emergent inverse metric,
two eigenvalues with opposite signs.  
For completeness, we will also discuss an extended 
4D toy-model calculation, which is slightly more
realistic as it allows for a direct interpolation
between the standard 4D Euclidean inverse metric
and the standard 4D Minkowski inverse metric.
Throughout this appendix, we use length units 
that set the IIB-matrix-model length scale to unity, $\ell=1$.

Both calculations start from the
multiple integral \eqref{eq:emergent-inverse-metric}
for spacetime dimension $D=2$ or $4$ at the spacetime point
\begin{subequations}\label{eq:appB-assumptions}
\begin{eqnarray} \label{eq:appB-assumptions-xmu-equals-zero}
x^{\mu} &=& 0 \,,
 \end{eqnarray}
with a simplified integrand having
\begin{eqnarray} \label{eq:appB-assumptions-rho-av-equals-1}
\langle\langle\, \rho(y)  \,\rangle\rangle&=& 1\,,
\\[1.0mm]
\label{eq:appB-assumptions-r-equals-1}
r(x,\,y)&=& 1\,,
\end{eqnarray}
and symmetric cutoffs on the integrals,
\begin{eqnarray} \label{eq:appB-assumptions-symmetric-cutoffs}
\int_{-1}^{1} dy^{0}\, \cdots \,\int_{-1}^{1} dy^{D-1}\,.
\end{eqnarray}
\end{subequations}
The only nontrivial contribution to the integrand
of Eq.~\eqref{eq:emergent-inverse-metric}
then comes from the correlation function  $f(x-y)$,
for which we will make two \emph{Ans\"{a}tze}.

%%\newpage%%tmp
\subsection{2D calculation}  
\label{appsub:2D-calculation}

For the first toy-model calculation, we take
\begin{subequations}\label{eq:D-is-2-ftest2}
\begin{eqnarray}
\label{eq:D-is-2}
D&=&2\,,
\\[1.0mm]
\label{eq:ftest2}
f_\text{test,2}(y)&=& \alpha + y^{0}\,y^{1}\,,
\end{eqnarray}
\end{subequations}
where the \textit{Ansatz} function \eqref{eq:ftest2}
combines a term that is even in both $y^{0}$ and $y^{1}$
with a term that is odd in both $y^{0}$ and $y^{1}$.
From Eq.~\eqref{eq:emergent-inverse-metric}
with simplifications \eqref{eq:appB-assumptions},
we then get the following multiple integral
for the emerging inverse metric:
\begin{equation}
\label{eq:inv-metric-test2-integrals}
g^{\mu\nu}_\text{test,2}(0) =
\int_{-1}^{1} dy^{0} \int_{-1}^{1} dy^{1}\; y^{\mu}\,y^{\nu}\;f_\text{test,2}(y)\,.
\end{equation}
The integrals are trivial and we obtain the inverse metric
\begin{subequations}\label{eq:inv-metric-test2-result-eigenval}
\begin{eqnarray} \label{eq:inv-metric-test2-result}
g^{\mu\nu}_{\alpha} (0)
&=&
\left(\begin{array}{cc}
    \;\;4\,\alpha/3\;\; & 4/9 \\[1mm]
    4/9 & \;\;4\,\alpha/3\;\;\\
  \end{array}\right)\,,
\end{eqnarray}
which has the following set of eigenvalues:
\begin{eqnarray} \label{eq:inv-metric-test2-eigenval}
\mathcal{E}_{\alpha} &=&
\frac{4}{9}\,
\Big\{\big(3\,\alpha-1 \big) ,\, \big(3\,\alpha+1 \big)\Big\}\,,
\end{eqnarray}
\end{subequations}

We now introduce an interpolation parameter $\rho$,
\begin{subequations}\label{eq:alpha-interpol}
\begin{eqnarray}
\alpha(\rho) &=& 1-\rho\,,
\\[1.0mm]
\rho &\in& [0,\,1]\,,
\end{eqnarray}
\end{subequations}
so that the inverse metric \eqref{eq:inv-metric-test2-result}
and its eigenvalues \eqref{eq:inv-metric-test2-eigenval} are given by
\begin{subequations}\label{eq:inv-metric-test2-result-eigenval-rho}
\begin{eqnarray} \label{eq:inv-metric-test2-result-rho}
g^{\mu\nu}_{\rho}(0)
&=&
\left(\begin{array}{cc}
    \;\;4\,(1-\rho)/3\;\; & 4/9 \\[1mm]
    4/9 & \;\;4\,(1-\rho)/3\;\; \\
  \end{array}\right)\,,
\\[1.0mm]
\label{eq:inv-metric-test2-eigenval-rho}
\mathcal{E}_{\rho} &=&
\frac{4}{9}\,
\Big\{\big(2-3\,\rho \big) ,\, \big(4-3\,\rho \big)\Big\}\,.
\end{eqnarray}
\end{subequations}
We see that we have obtained an inverse metric that interpolates
between a Euclidean signature for $\rho = 0$ and a Lorentzian signature
for $\rho = 1$,
\begin{subequations}\label{eq:inv-metric-test2-rho-0-and-1}
\begin{eqnarray} \label{eq:inv-metric-test2-rho-0}
\mathcal{E}_{\rho=0} &=& \Big\{ 8/9 ,\, 16/9 \Big\}\,,
\\[1.0mm]
\label{eq:inv-metric-test2-rho-1}
\mathcal{E}_{\rho=1} &=& \Big\{ -4/9 ,\, 4/9 \Big\}\,.
\end{eqnarray}
\end{subequations}
At $\rho=2/3$,
the inverse metric \eqref{eq:inv-metric-test2-result-eigenval-rho}
is degenerate, with a vanishing determinant.

%%\newpage%%tmp
The origin of the Lorentzian signature \eqref{eq:inv-metric-test2-rho-1}
is easy to understand. For $\rho=1$, the \textit{Ansatz}
parameter $\alpha=1-\rho$ in Eq.~\eqref{eq:ftest2} equals zero,
so that the integrand of Eq.~\eqref{eq:inv-metric-test2-integrals}
becomes simply $y^{\mu}\,y^{\nu}\;y^{0}\,y^{1}$.
The symmetric integrals \eqref{eq:inv-metric-test2-integrals} then 
vanish unless $\{\mu,\,\nu\}=\{0,\,1\}$ or $\{\mu,\,\nu\}=\{1,\,0\}$.
In other words, the matrix for the emergent inverse metric
\eqref{eq:inv-metric-test2-integrals}
is off-diagonal with entries $(2/3)^{2}=4/9$, so that the eigenvalues are
$\pm 4/9$. The off-diagonal matrix structure traces back to the
assumption that the correlation function $f(y)$,
for $\rho=1$ or $\alpha=0$, is given by a single monomial $y^{0}\,y^{1}$,
which is odd in both $y^{0}$ and $y^{1}$.

A final remark on this 2D calculation of a Lorentzian
signature is in order.
From the $\rho = 1$ inverse metric \eqref{eq:inv-metric-test2-result-rho},
we obtain, after a suitable coordinate transformation
(with $g^{\mu\nu} \to g^{\;\prime\;\mu\nu}$) and
a rescaling of $x^{0}$ and $x^{1}$ by an identical factor
(here, a factor $2/3$), the standard Minkowski form,
$g^{\;\prime\;\mu\nu}=\text{diag}(-1,\,1)$.
Instead of rescaling the coordinates, it is also possible,
for this simple case, to multiply the \textit{Ansatz}
function \eqref{eq:ftest2} by an appropriate overall factor
(here, a factor 9/4).

%%\newpage%%tmp
\subsection{4D calculation}  
\label{appsub:4D-calculation}

For the second toy-model calculation, we take
\begin{subequations}\label{eq:D-is-4-ftest4}
\begin{eqnarray}
\label{eq:D-is-4}
D&=& 4\,,
\\[1.0mm]
\label{eq:ftest4}
f_\text{test,4}(y)&=&
\alpha +\beta\,\Big[\big(y^{2}\big)^{2}+\big(y^{3}\big)^{2}\Big]
+ \gamma\,y^{0}\,y^{1}\,,
\end{eqnarray}
\end{subequations}
where the \textit{Ansatz} function \eqref{eq:ftest4}
combines two terms that are even in both $y^{0}$ and $y^{1}$
with one term that is odd in both $y^{0}$ and $y^{1}$.
From Eq.~\eqref{eq:emergent-inverse-metric}
with simplifications \eqref{eq:appB-assumptions},
we then get the emergent inverse metric
\begin{equation}
\label{eq:inv-metric-test4-integrals}
g^{\mu\nu}_\text{test,4}(0) =
\int_{-1}^{1} dy^{0} \int_{-1}^{1} dy^{1} \int_{-1}^{1} dy^{2} \int_{-1}^{1} dy^{3}
\; y^{\mu}\,y^{\nu}\;f_\text{test,4}(y)\,.
\end{equation}
Again, the integrals are trivial and we obtain
\begin{subequations}\label{eq:inv-metric-test4-result-eigenval}
\begin{eqnarray} \label{eq:inv-metric-test4-result}
\hspace*{-10mm}
g^{\mu\nu}_{\alpha\beta\gamma}(0)
&=&
\frac{16}{9}\,
\left(
  \begin{array}{cccc}
\;3\,\alpha+2\,\beta\;\ & \gamma & 0 & 0\\
\gamma & \;3\,\alpha+2\,\beta\; & 0 & 0\\
0 & 0 &  \;3\,\alpha+(14/5)\,\beta\; &  0\\
0 & 0 & 0 &\;3\,\alpha+(14/5)\,\beta\;\\
  \end{array}
\right),
\end{eqnarray}
which has the following set of eigenvalues:
\begin{eqnarray} \label{eq:inv-metric-test4-eigenval}
\hspace*{-10mm}
\mathcal{E}_{\alpha\beta\gamma} &=&
\frac{16}{9}\,\left\{
\Big( 3\,\alpha + 2\,\beta + \gamma \Big),\,
\Big( 3\,\alpha + 2\,\beta - \gamma \Big),\,
\left( 3\,\alpha + \frac{14}{5}\,\beta \right),\,
\left( 3\,\alpha + \frac{14}{5}\,\beta \right)
\right\}.
\end{eqnarray}
\end{subequations}

Let us now introduce an interpolation parameter $\sigma$,
\begin{subequations}\label{eq:alpha-beta-gamma-interpol}
\begin{eqnarray}
\alpha(\sigma)&=&\frac{3}{32}\,\big( 2-7\,\sigma  \big)\,,
\\[1.0mm]
\beta(\sigma) &=& \frac{45}{64}\,\sigma \,,
\\[1.0mm]
\gamma(\sigma) &=& -\frac{9}{16}\,\sigma \,,
\\[1.0mm]
\sigma &\in& [0,\,1]\,,
\end{eqnarray}
\end{subequations}
so that the inverse metric \eqref{eq:inv-metric-test4-result}
and its eigenvalues \eqref{eq:inv-metric-test4-eigenval}
are given by
\begin{subequations}\label{eq:inv-metric-test4-matrix-eigenval-sigma}
\begin{eqnarray}
\label{eq:inv-metric-test4-matrix-sigma}
g^{\mu\nu}_{\sigma}(0)
&=&
\left(
  \begin{array}{cccc}
    \;\;1 - \sigma\;\; & \;\;-\sigma\;\; & \;0\;\; & \;0\;\;\\
    -\sigma & \;\;1 - \sigma\;\; & 0 & 0 \\
    0 & 0 & 1 & 0 \\
    0 & 0 & 0 & 1 \\
  \end{array}
\right)\,,
\\[1.0mm]
\label{eq:inv-metric-test4-eigenval-sigma}
\mathcal{E}_{\sigma} &=&
\Big\{ 1-2\,\sigma,\,1,\, 1,\, 1  \Big\}\,.
\end{eqnarray}
\end{subequations}
From Eq.~\eqref{eq:inv-metric-test4-matrix-sigma} for  $\sigma=0$,
we immediately have the standard Euclidean inverse metric,
\begin{subequations}\label{eq:inv-metric-test4-result-sigma-0-1}
\begin{equation}
g^{\mu\nu}_{\sigma=0}(0)=\text{diag} \Big(1,\, 1,\, 1,\, 1\Big)\,,
\end{equation}
while, from Eq.~\eqref{eq:inv-metric-test4-matrix-sigma} for $\sigma=1$,
we obtain, after a suitable coordinate transformation
(with $g^{\mu\nu} \to g^{\;\prime\;\mu\nu}$),
the standard Minkowski inverse metric,
\begin{equation}
g^{\;\prime\;\mu\nu}_{\sigma=1}(0) = \text{diag} \Big(-1,\, 1,\, 1,\, 1\Big)\,.
\end{equation}
\end{subequations}
Again, we interpolate smoothly between a Euclidean signature
($\sigma=0$) and a Lorentzian signature ($\sigma=1$).
At $\sigma=1/2$,
the inverse metric \eqref{eq:inv-metric-test4-matrix-eigenval-sigma}
is degenerate, with a vanishing determinant.

The expression \eqref{eq:emergent-inverse-metric}
for the emergent inverse metric, first
proposed in Ref.~\cite{Aoki-etal-review-1999}
and reinterpreted in the present paper,
has the potential to give either a Euclidean or a Lorentzian
inverse metric, depending on the functional behavior of the correlation
functions $r(x,\,y)$ and $f(x-y)$, which result
from the detailed structure of the emerging spacetime points.
In principle, it is even possible to get a Lorentzian emergent
inverse metric from the Euclidean IIB matrix model, provided 
that the correlation functions have the appropriate structure
\big[a Euclidean toy-model calculation for $D=4$   %%XXX
may give the inverse metric \eqref{eq:inv-metric-test4-integrals}
with $y^{0}$ replaced by $y^{4}$ and $f_\text{test,4}(y)$ by
$f_\text{test,E4}(y)=1 - \gamma\,\big(y^{1}\, y^{2} + y^{1}\, y^{3} 
+ y^{1}\, y^{4} + y^{2}\, y^{3} + y^{2}\, y^{4} + y^{3}\, y^{4}\big)$,
and then finds the Lorentzian signature $(- + + + )$  
for parameter values $\gamma>1$\,\big].
This last observation, 
%%if true,    
if applicable,   %%XXX
would remove the
need for working with the 
%(difficult) 
(possibly more difficult)   %%XXX
Lorentzian IIB matrix model
and the first two of the five preliminary remarks in
Sect.~\ref{sec:Intro} would no longer apply.

\end{document}